\documentclass[fleqn,10pt]{wlscirep}
\usepackage{dcolumn}
\newcolumntype{C}[1]{>{\centering}m{#1}}
\newcommand{\etal}{\textit{et al.}}
\newcommand{\bv}[1]{{\boldsymbol #1}}

\title{Fermi surface and effective masses in photoemission response \\of the \boldmath (Ba$_{1-x}$K$_x$)Fe$_2$As$_2$ \unboldmath superconductor} 
\author[1,*]{Gerald Derondeau}   
\author[2]{Federico Bisti} 
\author[2,3]{Masaki Kobayashi} 
\author[1]{J\"urgen Braun} 
\author[1]{Hubert Ebert} 
\author[2]{Victor A. Rogalev} 
\author[2]{Ming Shi}
\author[2]{Thorsten~Schmitt} 
\author[2,4,5]{Junzhang Ma}
\author[4,5]{Hong Ding}
\author[2,$\dagger$]{Vladimir N. Strocov} 
\author[6,$\ddagger$]{J\'an Min\'ar}

\affil[1]{Department  Chemie,  Physikalische  Chemie,  Universit\"at  M\"unchen, Butenandtstr.  5-13, 81377 M\"unchen, Germany\\}
\affil[2]{Swiss Light Source, Paul Scherrer Institute, CH-5232 Villigen PSI, Switzerland\\}
\affil[3]{Department of Applied Chemistry, School of Engineering, University of Tokyo, 7-3-1 Hongo, Bunkyo-ku, Tokyo 113-8656, Japan\\}
\affil[4]{Beijing National Laboratory for Condensed Matter Physics, China \\}
\affil[5]{Institute of Physics, Chinese Academy of Sciences, Beijing 100190, China\\}
\affil[6]{NewTechnologies-Research Center, University of West Bohemia, Pilsen, Czech Republic\\}

\affil[*]{gerald.derondeau@cup.uni-muenchen.de}
\affil[$\dagger$]{vladimir.strocov@psi.ch}
\affil[$\ddagger$]{jminar@ntc.zcu.cz}

\date{\today}% It is always \today, today,
\begin{abstract}
The angle-resolved photoemission spectra of the superconductor
(Ba$_{1-x}$K$_x$)Fe$_2$As$_2$ have been investigated accounting
coherently for spin-orbit coupling, disorder and electron correlation
effects in the valence bands combined with final state, matrix element
and surface effects.  
Our results explain the previously obscured origins of all salient
features of the ARPES response of this paradigm pnictide compound and
reveal the origin of the Lifshitz transition. Comparison of calculated
ARPES spectra with the underlying DMFT band structure shows an
important impact of final state effects, which result for
three-dimensional states in a deviation of the ARPES spectra from the
true spectral function. In particular, the apparent effective mass
enhancement seen in the ARPES response is not an entirely intrinsic
property of the quasiparticle valence bands but may have a significant
extrinsic contribution from the photoemission process and thus differ
from its true value. Because this effect is more pronounced for low
photoexcitation energies, soft-X-ray ARPES delivers more accurate
values of the mass enhancement due to a sharp definition of the 3D
electron momentum. To demonstrate this effect in addition to the
theoretical study, we show here new state
of the art soft-X-ray and polarisation dependent ARPES measurments.  
\end{abstract}
\begin{document}
\flushbottom
\maketitle

\thispagestyle{empty}

\noindent                

%%%%%%%%%%%%%%%%%%%%%%%%%%%%%%%%%%%%%%%%%%%%%%%%%%%%%%%%%%%%%%%%%%%%%%%%
%%%%%%%%%%%%%%%%%%%%%%%%%%%%%%%%%%%%%%%%%%%%%%%%%%%%%%%%%%%%%%%%%%%%%%%%
\section*{Introduction}
%%%%%%%%%%%%%%%%%%%%%%%%%%%%%%%%%%%%%%%%%%%%%%%%%%%%%%%%%%%%%%%%%%%%%%%%
%%%%%%%%%%%%%%%%%%%%%%%%%%%%%%%%%%%%%%%%%%%%%%%%%%%%%%%%%%%%%%%%%%%%%%%%
The iron pnictides are nowadays one of the most studied examples for
unconventional superconductivity. Due to their complex properties
standard theoretical methods based on a local density approximation
(LDA) within density functional theory (DFT) often
fail.\cite{MJ09,MSJD08,MJB+08,PhysRevB.79.224511,PhysRevB.81.214503} This is especially true if one tries to
explain angle-resolved photoemission (ARPES) spectra of the iron
pnictides.\cite{ZIE+09,EIZ+09a,EKZ+11,EIZ+09,WCM+12,BLT+13,EZK+14,BLL+16}
In this context a significant discrepancy between the effective masses
derived from experimental ARPES spectra $m^*_\text{exp}$ and from LDA
band structure calculations $m^*_\text{LDA}$ was
reported.\cite{BLT+13,BLL+16,PhysRevB.93.205143} Correct trends in the effective masses
can be observed using dynamical a mean-field theory (DMFT) approach
which quantifies the importance of correlation effects for the iron
pnictides.\cite{TSK12,BGJV14} 

Various advanced approaches have been applied in the field, accounting
for different phenomena. This covers the treatment of disorder in an
appropriate way \cite{BLGK12,WBW+13,KJ14}, the inclusion of spin-orbit
coupling (SOC) \cite{BEL+16} and in order to calculate ARPES spectra
correctly the influence of matrix element effects and surface effects
was recently stressed \cite{WRH+12,DBEM16}. Finally, electron-electron
correlation effects are one of the most important issues
discussed.\cite{YHK11,ZLO+11,WCM+12,FJV12,TSK12,XRR+13,YHK14} All
these aspects were shown to play a crucial role for the iron
pnictides, yet most approaches so far can deal with only one of these
issues at the same time.

In this work we will present a theoretical approach which accounts for
all of the above mentioned issues leading in this way to a very
satisfactory agreement with experimental ARPES data of the iron
pnictides. Here we investigate one of the most prominent prototype
systems in the family of iron pnictides, namely the K substituted
(Ba$_{1-x}$K$_x$)Fe$_2$As$_2$ compound \cite{RTJ+08,RTJ08}, which was
extensively studied by
ARPES.\cite{ZIE+09,EIZ+09,EIZ+09a,EKZ+11,EZK+14,NSR+11,XRS+13} There
is common agreement, that the Fermi surface (FS) of this compound is
quite complex and cannot be obtained from plain DFT calculations. In
fact, an exceptional propeller-like FS topology at the $\bar{\mathrm
  X}$ point is found \cite{ZIE+09,EIZ+09a,EKZ+11} which is discussed
in terms of a Lifshitz transition, meaning topological changes in the
FS which mark the onset of superconductivity.\cite{LKF+10,KJ14} Also a
rather puzzling change in the intensity distribution at neighboring
$\bar \Gamma$ points is known.\cite{ZIE+09} Until now there is no
theoretical work which would explain all the salient features of the
ARPES spectra of (Ba$_{1-x}$K$_x$)Fe$_2$As$_2$. Our theoretical work
is supported by the new bulk sensitive ARPES measurments done 
in the soft X-ray regime using various linear polarisation of the light.

%%%%%%%%%%%%%%%%%%%%%%%%%%%%%%%%%%%%%%%%%%%%%%%%%%%%%%%%%%%%%%%%%%%%%%%%
%%%%%%%%%%%%%%%%%%%%%%%%%%%%%%%%%%%%%%%%%%%%%%%%%%%%%%%%%%%%%%%%%%%%%%%%
\section*{Results}
%%%%%%%%%%%%%%%%%%%%%%%%%%%%%%%%%%%%%%%%%%%%%%%%%%%%%%%%%%%%%%%%%%%%%%%%
%%%%%%%%%%%%%%%%%%%%%%%%%%%%%%%%%%%%%%%%%%%%%%%%%%%%%%%%%%%%%%%%%%%%%%%%

%%%%%%%%%%%%%%%%%%%%%%%%%%%%%%%%%%%%%%%%%%%%%%%%%%%%%%%%%%%%%%%%%%%%%%%%%%%%%
\begin{figure*}[t]
%%%%%%%%%%%%%%%%%%%%%%%%%%%%%%%%%%%%%%%%%%%%%%%%%%%%%%%%%%%%%%%%%%%%%%%%%%%%%
\scalebox{1.04}{\includegraphics[clip]{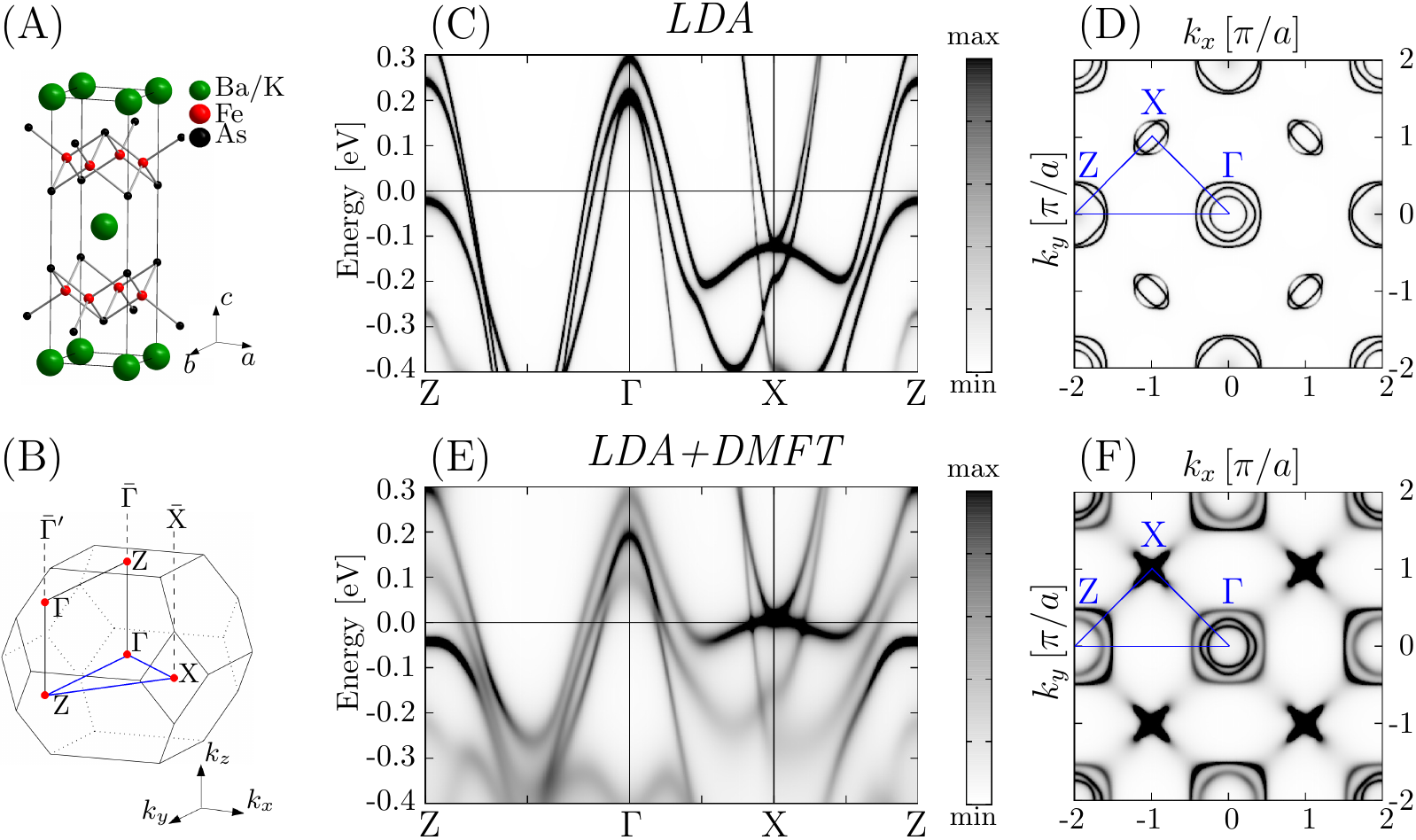}}
\caption{(A) Crystallographic unit cell of tetragonal (Ba$_{1-x}$K$_x$)Fe$_2$As$_2$ with (B) corresponding Brillouin zones indicating the important high symmetric points. $\bar \Gamma$, $\bar \Gamma '$ and $\bar{\mathrm{X}}$ indicate the two-dimensional Brillouin zone for a $(001)$ orientated surface.
(C + D) BSF and FS of (Ba$_{0.6}$K$_{0.4}$)Fe$_2$As$_2$ calculated on the basis of LDA. (E + F) Corresponding BSF and FS of (Ba$_{0.6}$K$_{0.4}$)Fe$_2$As$_2$ calculated on the basis of LDA+DMFT. The blue lines always indicate the path chosen for the presented band structure. }\label{Fig_1}
\end{figure*}
%%%%%%%%%%%%%%%%%%%%%%%%%%%%%%%%%%%%%%%%%%%%%%%%%%%%%%%%%%%%%%%%%%%%%%%%

\subsection*{Impact of correlations on the electronic structure}
%%%%%%%%%%%%%%%%%%%%%%%%%%%%%%%%%%%%%%%%%%%%%%%%%%%%%%%%%%%%%%%%%%%%%%%%

The crystal structure of (Ba$_{0.6}$K$_{0.4}$)Fe$_2$As$_2$ is shown in
Fig.~\ref{Fig_1}~(A), with the corresponding Brillouin zone~(BZ) and
its two-dimensional counterpart for a (001) orientated surface given
in Fig.~\ref{Fig_1}~(B). The electronic structure is represented by
means of the Bloch spectral function (BSF), which has the significant
advantage that in the presented approach all disorder effects induced
through substitution are fully accounted for (for more details see
review \cite{EKM11}). The LDA based band structure is shown in Fig.~\ref{Fig_1}~(C) with the corresponding Fermi surface (FS) cut shown in Fig.~\ref{Fig_1}~(D). The topology of this FS cut fails to explain the Fermi surface seen by ARPES.\cite{ZIE+09} It can neither reproduce the well-known propeller-like features at the $\bar{\mathrm X}$ point, nor it can explain the flower-like topology at $\bar \Gamma '$.\cite{ZIE+09} Thus, the applied LDA approach is insufficient to deal accurately with these prominent features. To account for the necessary correlation effects fully self-consistently, we have applied subsequent LDA+DMFT calculations. We used for Fe an averaged on-site Coulomb interaction $U=3.0$~eV and an exchange interaction $J=0.9$~eV, which are commonly used for the iron pnictides.\cite{WCM+12,FFVJ12,APV+09} The impact of correlation effects represented by the DMFT on the band structure in Fig.~\ref{Fig_1}~(E) and on the Fermi surface cut in Fig.~\ref{Fig_1}~(F) is tremendous. We see strong renormalization of the $d_{xy}$ and $d_{xz/yz}$ bands, in agreement with other literature.\cite{BGJV14} However, most prominent are the changes around the X point where a significant upwards shift of the bands towards the Fermi level ($E_F$) leads to the hole and electron pockets responsible for the appearance of the propeller like topology at $\bar{\mathrm{X}}$ in agreement with experimental ARPES data.\cite{ZIE+09,EIZ+09a,EKZ+11} 
Note, that a similar upwards shift at X of around 0.1\;eV was also observed by Werner \etal\cite{WCM+12},
although the qualitative agreement of our results with experiment
seems slightly better. The X point in our notation
corresponds to the M point in the notation used by Werner \etal \cite{WCM+12}. In comparison, Werner \etal~used a frequency
dependent screening which leads to strong incoherence. However, based
on suggestions by Tomczak \etal\cite{TSK12} and because we look only
at energies close to the Fermi level it seems like an acceptable
approximation to use a static coulomb interaction $U$. In particular,
we are able to fully account for the chemical disorder of the K-doped
compound in terms of the coherent potential approximation (CPA) which
seems to be more relevant for the problem at hand. Considering the LDA
band structure from Fig.~\ref{Fig_1}~(C) one can already see strong
band broadening for the hole band of interest at X. Thus, incoherence
due to disorder effects is strongest for explicitly this band and it
would be invalid to neglect this. Consequently, strong incoherence at
X for the LDA+DMFT band structure makes it difficult to resolve the
exact band shape. Based on the ARPES data of Zabolotnyy
\etal\cite{ZIE+09} the electron pocket and the hole pocket at X should
hybridize even more than in the presented calculations. This effect is
covered in our results due to the strong incoherence and it might be
also a shortcoming of the applied FLEX DMFT solver.  

Still, the consequence of this phenomenon is a topological change in the FS contour, indicating a so-called Lifshitz transition which is crucial for the emergence of superconductivity.\cite{KJ14,LKF+10} This Lifshitz transition was already discussed for high K concentrations ($x \approx 0.9$) within a LDA framework.\cite{XRS+13,KJ14} The present as well as previous experimental work \cite{ZIE+09,EIZ+09a,EKZ+11}, however, clearly shows the emergence of the petal topology around $\bar{\mathrm X}$ already for optimal doping ($x \approx 0.4$). Based on the latter discussion and by comparing Fig.~\ref{Fig_1} (C) and (D) one can see that the origin of the Lifshitz transition at lower K concentrations is fully controlled by correlation effects accounted for by the applied LDA+DMFT approach. Consequently, the Lifshitz transition can also qualitatively explain the breakdown of magnetic order for (Ba$_{1-x}$K$_x$)Fe$_2$As$_2$, which takes place at low doping ratios $x < 0.4$ \cite{RTJ08}, as it destroys the nesting condition \cite{KLH+12}. More details showing the clear dependence on the Coulomb interaction $U$ are found in the Supplemental Material.\cite{Supp} Although, the applied self-consistent DMFT approach has brought important new insights on the topology around $\bar{\mathrm X}$, it is not able to reproduce the flower-like intensity distribution observed at $\bar \Gamma '$ compared to $\bar \Gamma$.\cite{ZIE+09}\\ 

\subsection*{Impact of ARPES response effects}
%%%%%%%%%%%%%%%%%%%%%%%%%%%%%%%%%%%%%%%%%%%%%%%%%%%%%%%%%%%%%%%%%%%%%%%%

%%%%%%%%%%%%%%%%%%%%%%%%%%%%%%%%%%%%%%%%%%%%%%%%%%%%%%%%%%%%%%%%%%%%%%%%%%%%%
\begin{figure*}[ht]
%%%%%%%%%%%%%%%%%%%%%%%%%%%%%%%%%%%%%%%%%%%%%%%%%%%%%%%%%%%%%%%%%%%%%%%%%%%%%
\centering
\scalebox{1.07}{\includegraphics[clip]{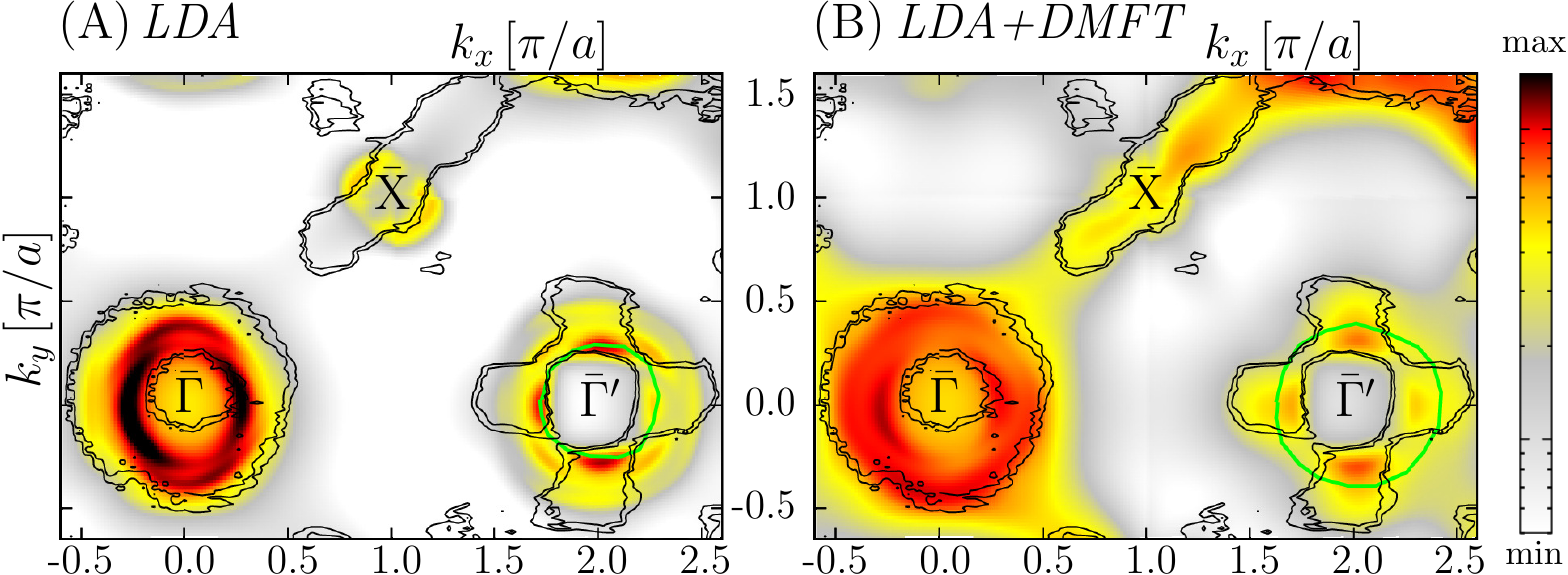}} 
\caption{Fermi surfaces cuts of (Ba$_{0.6}$K$_{0.4}$)Fe$_2$As$_2$ for $h\nu = 75$~eV as seen by one step model ARPES calculations for (A) LDA and (B) LDA+DMFT. The overlay of black isolines always corresponds to experimental ARPES data taken with permission from Zabolotnyy \etal \cite{ZIE+09}. The green solid lines are guides for the eyes to indicate surface state related features. }\label{Fig_2}
\end{figure*}
%%%%%%%%%%%%%%%%%%%%%%%%%%%%%%%%%%%%%%%%%%%%%%%%%%%%%%%%%%%%%%%%%%%%%%%%%%%%%

To understand this flower-like feature, additional calculations based
on the one-step model of ARPES have been done, accounting for the
experimental geometry \cite{SWS+14,SKW+14} including surface effects
as well as matrix element effects. These calculations were performed
using the LDA potentials and within the LDA+DMFT framework. The
corresponding spectroscopic Fermi surface cuts obtained from the
one-step model ARPES calculations of (Ba$_{0.6}$K$_{0.4}$)Fe$_2$As$_2$
for $h\nu = 75$~eV are shown in Fig.~\ref{Fig_2} (A) for LDA and in
Fig.~\ref{Fig_2} (B) for the LDA+DMFT calculations. For comparison we
show in both pictures the original experimental data from Zabolotnyy
\etal~\cite{ZIE+09} as an overlay of black isolines, measured at $h\nu
= 80$~eV. Please note that the theoretical photon energy of $h\nu =
75$~eV corresponds to the  experimental value of $h\nu
= 80$~eV and it has been found by the theoretical k$_z$ scan (not
shown here). This is typical shift in the energy of final time reversal
LEED state and it is  due to the limitations of the density functional
theory.    

The ARPES calculations based on the LDA+DMFT shown in Fig.~\ref{Fig_2}
(B) reveal very good agreement with the experimental data, concerning
the Lifshitz transition induced propeller structure at $\bar{\mathrm
  X}$ (only one part of the propeller is clearly visible for the
chosen light polarization as was also found in experiment
\cite{ZIE+09}) indicating that the previously discussed electronic
structure is correctly reproduced. Furthermore, we obtain good
agreement concerning the flower intensity distribution at $\bar \Gamma
'$. It should be stressed that there is no alternation of this circle
and flower topology around $\bar \Gamma$ and $\bar \Gamma'$,
respectively, with alternating $\Gamma$ and Z points in the $k_z$
direction by changing $h\nu$. Thus, the origin of this interesting
topology is not connected to the alternation between $\Gamma$ and Z
points of the bulk Brillouin zone.

The general appearance of two different shapes between $\bar \Gamma$ and $\bar \Gamma '$ can be explained by the structure factor and the light polarization in terms of a 1-Fe or 2-Fe cell, as discussed by e.g. Moreschini \etal~\cite{MLL+14} or Lin \etal~\cite{LBW+11}. Note, that a correct treatment of the phase difference between two atoms of a unit cell and the light polarization is by construction included in the one-step model of photoemission (see Ref. \cite{LKS+15}), thus the theory can sufficiently account for this. However, we find that at the same time other effects can contribute, in order to obtain this flower topology in agreement with experiment. As can be seen from Fig.~\ref{Fig_2}~(A) the intensity distribution at $\bar \Gamma'$ has a fourfold rotational symmetry, although the flower-like topology is not adequately reproduced compared to the result in Fig.~\ref{Fig_2}~(B). This difference cannot be explained without additional contributions.

It is known that the intensity distributions in ARPES might change for neighboring Brillouin zones between $\bar \Gamma$ and $\bar \Gamma'$ due to matrix element effects, however, such a strong change in the intensity distribution as seen in experiment \cite{ZIE+09} and reproduced in Fig.~\ref{Fig_2}~(B) is rather uncommon and unexpected. Yet, it is also known that the influence of matrix element effects can be enhanced in the vicinity of surface related states. 
Surface phenomena can be investigated by the applied method, as it is explained in more detail in Ref. \cite{BD02}. In particular, we have recently shown that surface states have a significant influence on the ARPES spectra of Co-doped BaFe$_2$As$_2$ \cite{DBEM16}.
Subsequently, one is able to identify in the spectra of (Ba$_{0.6}$K$_{0.4}$)Fe$_2$As$_2$ surface resonance states which wave functions have bulk Bloch asymptotic behavior and exhibit a strong resonance at the vicinity of the surface. This means such surface resonances can show a $k_z$ dispersion and they can be observed also for comparably high photon energies. The positions of these ring-shaped surface resonances is marked with solid green lines as an overlay in Fig.~\ref{Fig_2}. For LDA in Fig.~\ref{Fig_2} (A) one can see that this surface resonance is compressed and thus its influence on the intensity distribution at $\bar \Gamma'$ is less significant. In comparison, the surface resonance is shifted for the LDA+DMFT calculation in Fig.~\ref{Fig_2}~(B) where it cuts precisely through the clearly visible petals of the flower topology, affecting the intensity distribution at this position. We believe, that these contributions from the surface resonances can add up to the commonly discussed explanation based on the 1-Fe/2-Fe scheme, giving finally an overall good agreement with experimental data.

\subsection*{Bulk sensitive ARPES experiments}
%%%%%%%%%%%%%%%%%%%%%%%%%%%%%%%%%%%%%%%%%%%%%%%%%%%%%%%%%%%%%%%%%%%%%%%%

%%%%%%%%%%%%%%%%%%%%%%%%%%%%%%%%%%%%%%%%%%%%%%%%%%%%%%%%%%%%%%%%%%%%%%%%%%%%%
\begin{figure*}[ht]
%%%%%%%%%%%%%%%%%%%%%%%%%%%%%%%%%%%%%%%%%%%%%%%%%%%%%%%%%%%%%%%%%%%%%%%%%%%%%
\centering
\scalebox{1.07}{\includegraphics[clip]{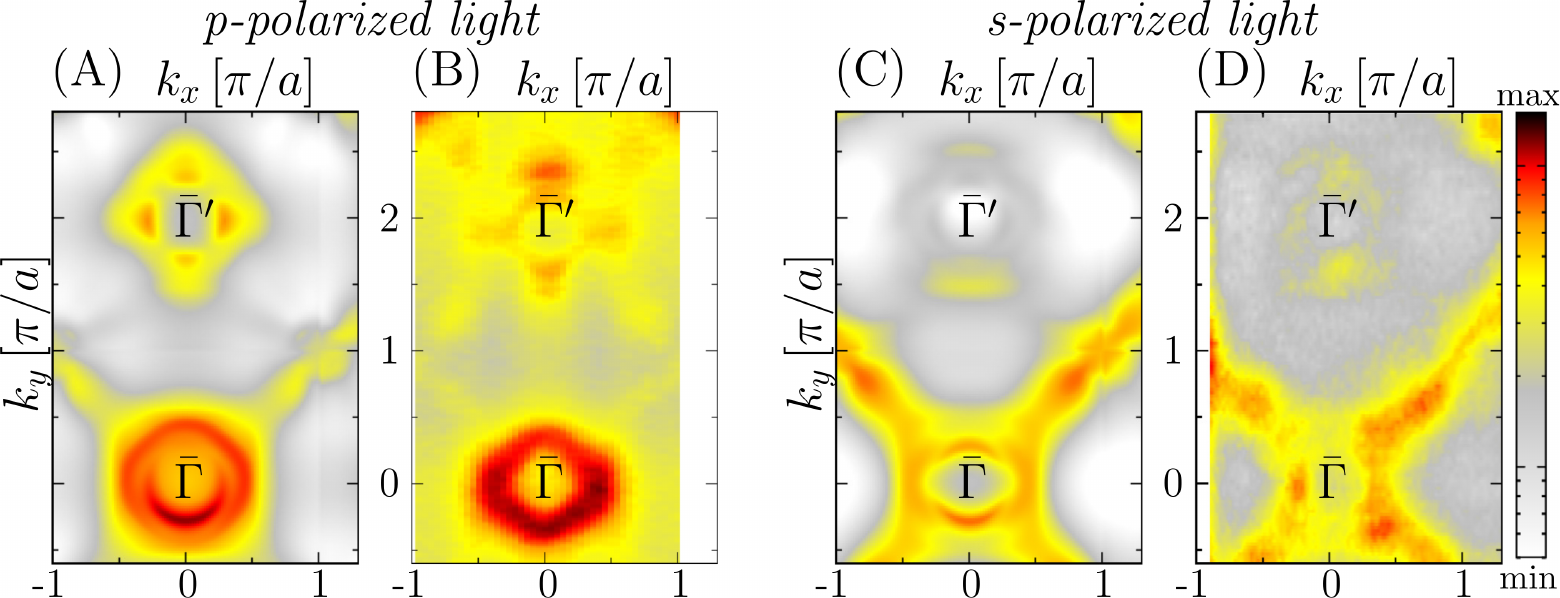}} 
\caption{Fermi surface cuts of (Ba$_{0.6}$K$_{0.4}$)Fe$_2$As$_2$ for (A+C) $h\nu = 425$~eV ARPES calculation
using LDA+DMFT and (B+D) $h\nu = 430$~eV experimental data. The incoming light was either (A+B) $p$-polarized
or (C+D) $s$-polarized. }\label{Fig_3}
\end{figure*}
%%%%%%%%%%%%%%%%%%%%%%%%%%%%%%%%%%%%%%%%%%%%%%%%%%%%%%%%%%%%%%%%%%%%%%%%%%%%%

Additional bulk sensitive soft-X-ray photoemission measurements for
$h\nu = 430$~eV were performed for samples of
(Ba$_{0.6}$K$_{0.4}$)Fe$_2$As$_2$.  In the first surface Brillouin zone, this photon energy sets k$_z$ to the $\Gamma$ point of the bulk one.
The resulting spectra are presented in Figs.~\ref{Fig_3} (B) and (D) for $p$-polarized and $s$-polarized light, respectively. Notably, for $p$-polarized light the flower shaped topology at $\bar \Gamma'$ is enhanced in intensity while for $s$-polarized light the propeller topologies at $\bar{\mathrm{X}}$ are enhanced. Corresponding calculations for $h\nu = 425$~eV are presented in Figs.~\ref{Fig_3} (A) and (C), which show very good agreement with the experimental data concerning the relevant topologies and the polarization dependence. 
Thus, these experiments are fully in line with the argumentation of this work so far and further validate our results.
Additional extended Fermi surface cuts for higher Brillouin zones can be found in the Supplemental Material.\cite{Supp}

\begin{figure*}[ht]
%%%%%%%%%%%%%%%%%%%%%%%%%%%%%%%%%%%%%%%%%%%%%%%%%%%%%%%%%%%%%%%%%%%%%%%%%%%%%
\centering
\scalebox{1.0}{\includegraphics[clip]{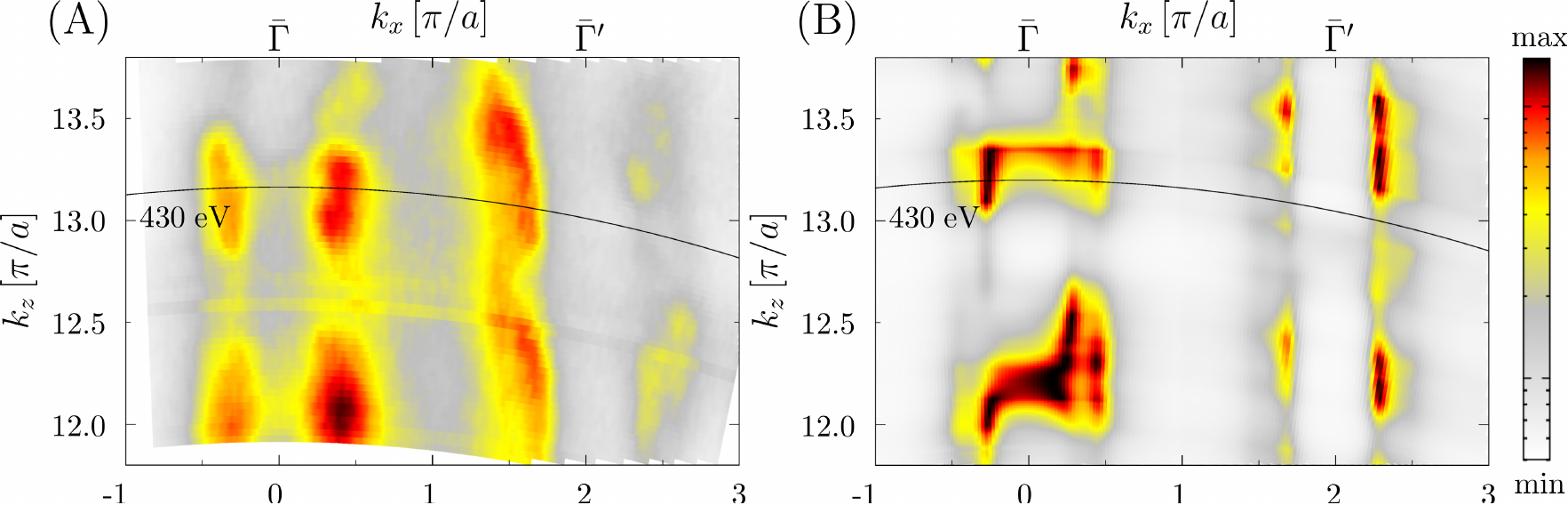}} 
\caption{Experimental (A) and theoretical (B)  $k_z$ scans with
  $p$-polarized light for (Ba$_{0.6}$K$_{0.4}$)Fe$_2$As$_2$. Clearly visible is the $k_z$ dispersion at
  $\bar \Gamma$ and $\bar \Gamma '$. The photon energy of $h\nu = 430$~eV and $h\nu = 392$~eV are marked with a black line. }\label{Fig_4}
\end{figure*}

Furthermore, the experimental and theoretical $k_z$ scan are shown in Fig.~\ref{Fig_4} with the photon energy of $h\nu = 430$~eV explicitly marked with the black line. This shows clearly the $k_z$ dispersion and thus the strong 3D character of the iron pnictides.

%%%%%%%%%%%%%%%%%%%%%%%%%%%%%%%%%%%%%%%%%%%%%%%%%%%%%%%%%%%%%%%%%%%%%%%%%%%%%
%%%%%%%%%%%%%%%%%%%%%%%%%%%%%%%%%%%%%%%%%%%%%%%%%%%%%%%%%%%%%%%%%%%%%%%%%%%%%

\subsection*{Effective masses derived from ARPES spectra}
%%%%%%%%%%%%%%%%%%%%%%%%%%%%%%%%%%%%%%%%%%%%%%%%%%%%%%%%%%%%%%%%%%%%%%%%

Finally, the electronic structure derived from BSF calculations and ARPES calculations was used to analyze the effective masses which are of great actual interest for the iron pnictides.\cite{TSK12,BLT+13,BGJV14,BLL+16} The results for the mass ratios of the inner and outer hole pockets around $\Gamma$ in the $\Gamma$Z direction and additionally for the hole pocket at X in the $\Gamma$X direction are summarized in Tab.~\ref{Tab_Mass}. More details are found in the Supplemental Material.\cite{Supp} 

\begin{table}[ht]
\renewcommand{\arraystretch}{1.1}
\centering
	\begin{tabular}{C{2.2cm}|C{1.3cm}|C{1.3cm}|C{1.3cm}}
	{$\frac{m^*_\text{DMFT}}{m^*_\text{LDA}}$} & {BSF}&{ARPES 75 eV}&
        {ARPES 425 eV}\tabularnewline 
	\hline	
	Inner pocket $\Gamma$ & 2.59 & 3.66 &  2.94 \tabularnewline 
	Outer pocket $\Gamma$ & 1.70 & 2.28 &  1.98 \tabularnewline 
        \hline
        Hole pocket X& 1.43 & 1.56 &  1.58 \tabularnewline
	\end{tabular}
\caption{Ratio of effective masses of $m^*_\text{DMFT}$ to
  $m^*_\text{LDA}$ for the Bloch spectral function (BSF) ground state calculations as well as for the ARPES calculations with $h\nu = 75$ eV and 425 eV, respectively. The values correspond to the inner and outer hole pockets around $\Gamma$ showing strong $k_z$ dispersion and to the hole pocket at X showing weak $k_z$ dispersion.\cite{Supp}} \label{Tab_Mass}
\end{table}

As it is commonly done, all values of $m^*$ are normalized to the LDA value $m^*_\text{LDA}$ deduced from the ground state BSF. First, we consider the mass enhancement at $\Gamma$ only, where the influence of DMFT on the band dispersion can be seen for the BSF, with an average mass enhancement of 2.15 (meaning an average over inner and outer hole pocket $\Gamma$), being in good agreement with literature (e.g. 2.04 for KFe$_2$As$_2$ \cite{BGJV14}). 
Of more interest is the apparent mass enhancement deduced from the ARPES calculations compared to the BSF band dispersion. The difference is attributed to the fact that the calculated ARPES spectra include not only the correlation effects of DMFT but also final state effects which, as explained below, modify the ARPES spectral shape. On the experimental side, such an apparent mass enhancement has already been observed in ARPES for e.g. the BaFe$_2$As$_2$ parent compound and connected with the $k_z$ dispersion of the valence states.\cite{BLT+13} The apparent mass enhancement is given by the fact that the ARPES response of the 3D valence states is formed by averaging of their matrix-element weighted $k_z$ dispersion over an interval of the intrinsic final state $k_z$ broadening ($\Delta k_z$) determined by the photoelectron mean free path $\lambda$.\cite{Str03,KBM07} 

As illustrated in Fig. S5 in the Supplemental Material \cite{Supp}, near the extremes of the valence band $k_z$ dispersions this averaging effectively shifts the ARPES peaks from true $k_z$ dispersions into the band interior (for detailed physical picture see Ref. \cite{Str03}). In $\bv k_{||}$ dependent ARPES intensities this shift is seen as an apparent bandwidth reduction and corresponding mass enhancement. One can expect a stronger influence of these final state effects at lower photon energies where $\Delta k_z$ is larger due to a smaller $\lambda$. Indeed, our calculations find significant differences in the mass enhancement at $\Gamma$ depending on $h\nu$. For the low $h\nu$ of 75 eV ($\Delta k_z = 0.2779$ \AA$^{-1}$, which makes about 30\% of the perpendicular BZ dimension) we find an average mass enhancement of 2.97 at $\Gamma$ which is higher than the value of 2.15 obtained from the BSF. The significantly higher $h\nu$ of 425~eV ($\Delta k_z =  0.1228$ \AA$^{-1}$) increases $\lambda$ and concomitantly improves the $k_z$ definition. The final state effects have therefore a less pronounced contribution, reducing the average mass enhancement at $\Gamma$ to 2.46. This is true for almost all bands in the iron pnictides as they are 3D materials with most bands showing a clear $k_z$ dispersion. One of the rare exceptions for the (Ba$_{1-x}$K$_x$)Fe$_2$As$_2$ compound is the hole pocket at X which has almost 2D character and shows hardly a $k_z$ dispersion as can be seen in Fig.~\ref{Fig_1} (C + E) for the path $\Gamma$XZ. In such a case one would expect significantly less influence of the final-state effects and indeed, Tab.~\ref{Tab_Mass} shows that the apparent mass enhancement for high and low $h\nu$ at X is almost the same and very similar to the BSF mass enhancement.
This finally explains discrepancies in the observed mass enhancement for the iron pnictides. To reduce these $h\nu$ dependent deviations of the ARPES response from the true 3D valence bands, we justify the use of higher $h\nu$ in the soft-X-ray regime to improve the $k_z$ definition. 

%%%%%%%%%%%%%%%%%%%%%%%%%%%%%%%%%%%%%%%%%%%%%%%%%%%%%%%%%%%%%%%%%%%%%%%%%%%%%
\begin{figure*}[ht]
%%%%%%%%%%%%%%%%%%%%%%%%%%%%%%%%%%%%%%%%%%%%%%%%%%%%%%%%%%%%%%%%%%%%%%%%%%%%%
\centering
\scalebox{1.0}{\includegraphics[clip]{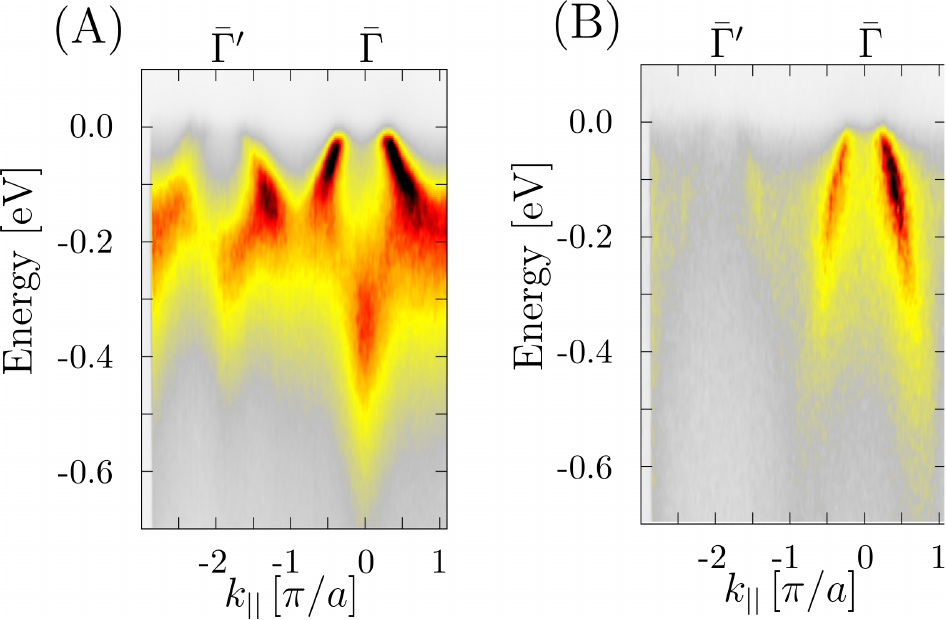}} 
\caption{Experimental band dispersions of
  (Ba$_{0.6}$K$_{0.4}$)Fe$_2$As$_2$ along the $\Gamma Z$ line of the Brillouin zone measured at $h\nu$ = 425 eV with (A) p-polarization, and (B) s-polarization selecting, respectively, the symmetric and antisymmetric d-states. }\label{Fig_5}
\end{figure*}
%%%%%%%%%%%%%%%%%%%%%%%%%%%%%%%%%%%%%%%%%%%%%%%%%%%%%%%%%%%%%%%%%%%%%%%%%%%%%
Finally, Fig.~\ref{Fig_5} shows the experimental band dispersions
measured at $h\nu= 425$~eV which correspond to $\Gamma Z$ direction in the bulk Brillouin
zone. In the soft-X-ray ARPES experimental geometry\cite{SWS+14,SKW+14} s-polarized
light excites the dxz, dx$^2$-y$^2$ and dz$^2$ states symmetric relative to the
$\Gamma Z M$ plane of the Brillouin zone, and p-polarized light the dyz and dxy
states antisymmetric relative to $\Gamma Z M$ \cite{Hajiri12}. On top of these dipole
selection rules, relative intensities of these bands alters between
the  $\Gamma$ and Z points because of the matrix element effects discussed
above.  
\begin{table}[ht]
\renewcommand{\arraystretch}{1.1}
\centering
	\begin{tabular}{C{2.2cm}|C{1.3cm}|C{1.3cm}|C{1.3cm}}
	&$\frac{m^*_\text{EXP}}{m^*_\text{LDA}}$& 
         $\frac{m^*_\text{EXP}}{m^*_\text{DMFT}}$&$\frac{m^*_\text{EXP}}{m^*_\text{ARPES DMFT}}$\tabularnewline 
	\hline	
	Inner pocket $\Gamma$  & 3.50 &1.35 &1.19\tabularnewline 
	Outer pocket $\Gamma$  & 2.26 &1.56&1.41\tabularnewline 
        \hline
	\end{tabular}
\caption{Ratio of effective masses of the experimental data with
  respect to the various theoretical values. Experimental data are
  taken from Fig. \ref{Fig_5}.  Values for $m^*_\text{DMFT}$ and 
  $m^*_\text{LDA}$ are taken from the ground state
  BSF and  $m^*_\text{ARPES DMFT}$  corresponds to the calculated one
  step model of photoemission spectra at $h\nu=$425eV as presented in
  Tab.~\ref{Tab_Mass}.} \label{Tab_Mass2}
\end{table}
From Fig.~\ref{Fig_5} we have extracted the corresponding experimental values of the effective
masses for the inner
and outer pockets at the $\Gamma$ point. In the Tab.~\ref{Tab_Mass2} we
summarize the results for band renormalization between the
experimental data and the corresponding ground state LDA and DMFT  
bands. Here we obtained an average value of
$\frac{m^*_\text{EXP}}{m^*_\text{LDA}}$=3.05 which, as we discussed
before, due to reduced value of k$_z$ broadening in the soft X-ray
regime can be taken as a reference value for future studies. As a last point we show
in the third column of Tab.~\ref{Tab_Mass2} the 
ratio of the experimental value ${m^*_\text{EXP}}$ to the theoretical
value ${m^*_\text{ARPES DMFT}}$ obtained from the one-step model ARPES calculations.
In the ideal case one would expect to reach a value of one. Due to the approximations we used for the DMFT
calculations, we obtain an average value of $1.3$. This discrepancy is
mainly due to the use of perturbative nature of the DMFT solver (FLEX), and as it was
discussed recently by  Werner et al. \cite{WCM+12} additional
dynamical screening effects are missing in our approach. However, by
comparing the spectra calculated within one step model of photoemission
with the state of the art experimental data, we are able to give a quantitative
measure of the theoretical approximations used here.
This would not be possible if we had used only ground state
calculations as demonstrated in the second column of Tab.~\ref{Tab_Mass2}.

%%%%%%%%%%%%%%%%%%%%%%%%%%%%%%%%%%%%%%%%%%%%%%%%%%%%%%%%%%
%%%%%%%%%%%%%%%%%%%%%%%%%%%%%%%%%%%%%%%%%%%%%%%%%%%%%%%%%%%%%%%%%%%%%%%%
\section*{Conclusions}
%%%%%%%%%%%%%%%%%%%%%%%%%%%%%%%%%%%%%%%%%%%%%%%%%%%%%%%%%%%%%%%%%%%%%%%%
%%%%%%%%%%%%%%%%%%%%%%%%%%%%%%%%%%%%%%%%%%%%%%%%%%%%%%%%%%%%%%%%%%%%%%%%

In conclusion, the presented LDA+DMFT+ARPES study is the first that quantitatively matches the theoretical description with the
experimental ARPES data on the paradigm high-temperature superconductor (Ba$_{0.6}$K$_{0.4}$)Fe$_2$As$_2$.
These results enables a better physical understanding of the unconventional superconductivity in pnictides and will be of
great importance for future studies on similar systems. In particular, the origin of the Lifshitz transition in
(Ba$_{1-x}$K$_x$)Fe$_2$As$_2$, crucial for its superconductivity, is identified as fully controlled by electron correlation
effects.
Furthermore, we have shown that due to the inherently 3D nature of the iron pnictides their ARPES response is significantly
influenced by final state effects, shifting the spectral peaks from the true quasiparticle valence bands. Their mass
enhancement apparent in the ARPES spectra is then different from the true value and, moreover, will depend on the photon
energy. Thus, the mass renormalization observed in previous ARPES works on iron pnictides is not an entirely intrinsic
property of the quasiparticle valence band structure or spectral function, but has a significant contribution due to a
peculiarity of the photoemission process extrinsic to the true valence band properties.

%%%%%%%%%%%%%%%%%%%%%%%%%%%%%%%%%%%%%%%%%%%%%%%%%%%%%%%%%%%%%%%%%%%%%%%%
%%%%%%%%%%%%%%%%%%%%%%%%%%%%%%%%%%%%%%%%%%%%%%%%%%%%%%%%%%%%%%%%%%%%%%%%
\section*{Methods}
%%%%%%%%%%%%%%%%%%%%%%%%%%%%%%%%%%%%%%%%%%%%%%%%%%%%%%%%%%%%%%%%%%%%%%%%
%%%%%%%%%%%%%%%%%%%%%%%%%%%%%%%%%%%%%%%%%%%%%%%%%%%%%%%%%%%%%%%%%%%%%%%%

\subsection*{Computational method}
%%%%%%%%%%%%%%%%%%%%%%%%%%%%%%%%%%%%%%%%%%%%%%%%%%%%%%%%%%%%%%%%%%%%%%%%

Within the present work, the multiple scattering Korringa-Kohn-Rostoker-Green fuction (KKR-GF) method was applied which allows to deal simultaneously with all mentioned spectroscopic and many-body aspects. All calculations have been performed within the fully relativistic four component Dirac formalism \cite{EKM11,SPR-KKR6.3_2}, accounting this way for all effects induced by spin-orbit coupling. Disorder effects are dealt with by means of the coherent potential approximation (CPA).\cite{DPM+14,DBEM16,KJ14} ARPES calculations are based on the one-step model of photoemission in its spin density matrix formulation using the experimental geometry.\cite{MBE13,BMK+14} Thus, the theory accounts for effects induced by the light polarization, matrix-element effects, final state effects and surface effects. To account for correlation effects fully self-consistently (concerning charge as well as self energy) the LDA+DMFT method using a FLEX solver was applied.\cite{Min11} For Fe an averaged on-site Coulomb interaction $U=3.0$~eV and an exchange interaction $J=0.9$~eV were applied. In the Supplemental Material calculations for different values of $U$ are shown.\cite{Supp}
The lattice constants of the tetragonal cell of (Ba$_{0.6}$K$_{0.4}$)Fe$_2$As$_2$ were taken from experimental data. \cite{RTJ08}

\subsection*{ARPES experiments}
%%%%%%%%%%%%%%%%%%%%%%%%%%%%%%%%%%%%%%%%%%%%%%%%%%%%%%%%%%%%%%%%%%%%%%%%

New ARPES experiments in the soft-X-ray photon energy ($h\nu$) range above 400\;eV were performed at the ADRESS beamline of the Swiss Light Source synchrotron facility.\cite{SWS+14,SKW+14} By using higher $h\nu$ compared to the conventional ultraviolet ARPES, higher bulk sensitivity is achieved due to an increase of the photoelectron mean free path $\lambda$ as expected from the well-known ``universal curve''. Crucial for 3D materials like the iron pnictides is that the increase of $\lambda$ results, by the Heisenberg uncertainty principle, in a sharp intrinsic definition of the momentum $k_z$ perpendicular to the surface.\cite{Str03} As explained in the paper, the latter becomes important for the correct evaluation of the true valence band dispersions and effective masses.

In particular, bulk sensitive soft-X-ray photoemission measurements
for $h\nu = 425$~eV were performed for in-situ cleaved samples of
(Ba$_{0.6}$K$_{0.4}$)Fe$_2$As$_2$ at a temperature of around 12\;K and
with an overall energy resolution of around 70~meV. 

%%%%%%%%%%%%%%%%%%%%%%%%%%%%%%%%%%%%%%%%%%%%%%%%%%%%%%%%%%%%%%%%%%%%%%%%
%%%%%%%%%%%%%%%%%%%%%%%%%%%%%%%%%%%%%%%%%%%%%%%%%%%%%%%%%%%%%%%%%%%%%%%%
%\section*{References}
%%%%%%%%%%%%%%%%%%%%%%%%%%%%%%%%%%%%%%%%%%%%%%%%%%%%%%%%%%%%%%%%%%%%%%%%
%%%%%%%%%%%%%%%%%%%%%%%%%%%%%%%%%%%%%%%%%%%%%%%%%%%%%%%%%%%%%%%%%%%%%%%%

%\bibliography{scirep}
%\bibliographystyle{aipnum.bst}

%%%%%%%%%%%%%%%%%%%%%%%%%%%%%%%%%%%%%%%%%%%%%%%%%%%%%%%%%%%%%%%%%%%%%%%%
%%%%%%%%%%%%%%%%%%%%%%%%%%%%%%%%%%%%%%%%%%%%%%%%%%%%%%%%%%%%%%%%%%%%%%%%
\section*{Acknowledgments}
%%%%%%%%%%%%%%%%%%%%%%%%%%%%%%%%%%%%%%%%%%%%%%%%%%%%%%%%%%%%%%%%%%%%%%%%
%%%%%%%%%%%%%%%%%%%%%%%%%%%%%%%%%%%%%%%%%%%%%%%%%%%%%%%%%%%%%%%%%%%%%%%%

We thank V. Zabolotnyy and S. Borisenko for allowing us to use their
experimental data. Special thanks goes to S. Biermann for discussions
and ideas. We acknowledge financial support from the  Deutsche
Forschungsgemeinschaft DFG (projects FOR 1346) and from the
Bundesministerium f\"ur Bildung und Forschung BMBF (project
05K16WMA). We further thank for the support from CENTEM PLUS
(L01402)  and CEDAMNF (CZ.02.1.01/0.0/0.0/15$\_$003/0000358) of Czech ministerium
MSMT. F. Bisti acknowledges the funding from the Swiss National 
Science Foundation under the grant agreement n.200021\_146890 and
European Community's Seventh Framework Programme (FP7/2007-2013) under
the grant agreement n.290605 (PSI-FELLOW/COFUND).

%%%%%%%%%%%%%%%%%%%%%%%%%%%%%%%%%%%%%%%%%%%%%%%%%%%%%%%%%%%%%%%%%%%%%%%%
%%%%%%%%%%%%%%%%%%%%%%%%%%%%%%%%%%%%%%%%%%%%%%%%%%%%%%%%%%%%%%%%%%%%%%%%
\section*{Author contributions statement}
%%%%%%%%%%%%%%%%%%%%%%%%%%%%%%%%%%%%%%%%%%%%%%%%%%%%%%%%%%%%%%%%%%%%%%%%
%%%%%%%%%%%%%%%%%%%%%%%%%%%%%%%%%%%%%%%%%%%%%%%%%%%%%%%%%%%%%%%%%%%%%%%%

G.D., V.N.S. and J.Mi. wrote the manuscript. F.B. V.A.R., M.K., T.S., J.Ma and V.N.S. performed the experiments and analyzed the data. G.D. performed the calculations. G.D., F.B., J.B., M.S., H.D., H.E., V.N.S. and J.Mi. participated at the discussions. J.Mi. supervised the theoretical part, V.N.S. supervised the experimental part.\\

%%%%%%%%%%%%%%%%%%%%%%%%%%%%%%%%%%%%%%%%%%%%%%%%%%%%%%%%%%%%%%%%%%%%%%%%
%%%%%%%%%%%%%%%%%%%%%%%%%%%%%%%%%%%%%%%%%%%%%%%%%%%%%%%%%%%%%%%%%%%%%%%%
\section*{Additional information}
%%%%%%%%%%%%%%%%%%%%%%%%%%%%%%%%%%%%%%%%%%%%%%%%%%%%%%%%%%%%%%%%%%%%%%%% 
%%%%%%%%%%%%%%%%%%%%%%%%%%%%%%%%%%%%%%%%%%%%%%%%%%%%%%%%%%%%%%%%%%%%%%%%

\noindent\textbf{Supplemental Material} accompanies this paper.\\

\noindent\textbf{The author(s) declare no competing financial interests.}

\end{document}